\documentclass[twocolumn,prd,amsfonts,showpacs,epsfig,floats,superscriptaddress,nofootinbib,floatfix,a4paper]{revtex4}
\usepackage{graphicx}% Include figure files
\usepackage{dcolumn}% Align table columns on decimal point
\usepackage{bm}% bold math
\usepackage{longtable}
\usepackage{amsmath}
\usepackage{mathrsfs}
\usepackage{amsfonts}

\usepackage{epstopdf}
\graphicspath{{./images/}}

\newcommand{\be} {\begin{equation}}
\newcommand{\ee} {\end{equation}}

\newcommand{\bea}{\begin{eqnarray}}
\newcommand{\eea}{\end{eqnarray}}
\newcommand{\bdm}{\begin{displaymath}}
\newcommand{\edm}{\end{displaymath}}
\newcommand{\ba} {\begin{array}}
\newcommand{\ea} {\end{array}}

\newcommand{\bfg}  {\begin{figure}}
\newcommand{\efg}  {\end{figure}}
\newcommand{\incgr} {\includegraphics}

\newcommand{\mbf}{\mathbf}
\newcommand{\tbf}{\textbf}

\def\eff {\rm eff}

\begin{document}

\bibliographystyle{apsrev}

\title{Affine parameterization of the dark sector: costraints from  WMAP5 and SDSS}

\author{Davide Pietrobon}
\affiliation{Institute of Cosmology and Gravitation, University
of Portsmouth, Mercantile House, Portsmouth PO1 2EG, United Kingdom}
\affiliation{Dipartimento di Fisica, Universit\`a di Roma ``Tor Vergata'',
via della Ricerca Scientifica 1, 00133 Roma, Italy}
\author{Amedeo Balbi}
\affiliation{Dipartimento di Fisica, Universit\`a di Roma ``Tor Vergata'',
via della Ricerca Scientifica 1, 00133 Roma, Italy}
\affiliation{INFN Sezione di Roma ``Tor Vergata'',
via della Ricerca Scientifica 1, 00133 Roma, Italy}
\author{Marco Bruni}
\affiliation{Institute of Cosmology and Gravitation, University
of Portsmouth, Mercantile House, Portsmouth PO1 2EG, United Kingdom}
\affiliation{Dipartimento di Fisica, Universit\`a di Roma ``Tor Vergata'',
via della Ricerca Scientifica 1, 00133 Roma, Italy}
\author{Claudia Quercellini}
\affiliation{Dipartimento di Fisica, Universit\`a di Roma ``Tor Vergata'',
via della Ricerca Scientifica 1, 00133 Roma, Italy}

\date{\today}

\begin{abstract}
We study a set of universe models where the dark sector  is described by  a perfect fluid with  an affine equation of  state $P=P_0+\alpha \rho$, focusing specifically on cosmological perturbations in a flat universe. We perform a Monte Carlo Markov Chain analysis  spanning the full parameter space of the model using the WMAP 5 years data and the SDSS LRG4
survey. The affine fluid can either  play the role of a unified dark matter (UDM), accounting for both dark matter and a cosmological constant, or work alongside cold dark matter (CDM), as a form of dark energy. A key ingredient is the sound speed, that depends on the nature of the fluid and that, for any given background model, adds a degree of freedom to the perturbations: in the
barotropic case the square of the sound speed is simply equal to the
affine parameter $\alpha$; if entropic perturbations are present the effective sound speed has to be specified as an additional parameter. In addition to the barotropic case, we consider the two limiting cases of
effective sound speed equal to 0 or 1. For $\alpha=c_s^2=0$ our UDM model is equivalent to the standard $\Lambda$CDM  with adiabatic perturbations. Apart of a trivial subcase, all models considered  satisfy the data constraints, with quite standard values for the usual cosmological parameters. In general our analysis confirms that cosmological datasets require both a collisionless massive and cold component to form the potential wells that lead to
structure formation, and an effective cosmological constant that drives the late accelerated expansion.
\end{abstract}

\maketitle

%------------------------------------------------------------------------------
\section{Introduction}
In the last few years cosmological observations have become increasingly accurate, allowing various models to be tested or even ruled out. The one that currently seems to satisfy most 
observational requirements is the so-called concordance or $\Lambda$CDM model \cite{Tegmark:2004,Spergel:2003}: a flat universe, with $\sim 4\%$ of baryons,  $\sim20\%$ of an unknown weakly-interacting heavy component (or dark matter), leaving the remaining $\sim 76\%$ in the form of a cosmological constant (or vacuum energy density) responsible for the late time acceleration of the universe  \cite{Riess:astro-ph/0611572,Perlmutter:astro-ph/9812133}. However, despite its simplicity, this model lacks solid theoretical motivations and actually seems to require \emph{ad hoc} assumptions, both on dark matter and the cosmological constant. Many hypotheses have been proposed to alleviate the problem of the cosmological constant and the related one of why the density of the two unknown components are of the same order of magnitude in the present universe (the so-called coincidence problem). For an incomplete list see \cite{bento03,Kamenshchik00,2008arXiv0807.1020B,
sahniwang00,mainini05,giannakis05,beca05,chimento04,scherrer04} and the recent review on dark energy \cite{DErevCopeland07}.

Observationally, the dark sector is degenerate: by definition, dark components can be probed only through gravitational effects, leaving open a wide range of possibilities regarding their nature and possible interactions \cite{2007astro.ph..2615K}. In this paper we investigate an effective model for the dark sector, based on the affine parameterisation of the equation of state \cite{Ananda:astro-ph/0512224}:
\begin{equation}
\label{affine_pressure}
p_{X}=P_0+\alpha\rho_{X},
\end{equation}
where $p_X$ is the pressure, $\rho_X$ is the energy density, and $P_0$ and $\alpha$ are constant parameters; this leads to a time dependent equation of state parameter
\begin{equation}
\label{w_eq}
w_{X}=\frac{P_0}{\rho_{X}}+\alpha.
\end{equation}
An interesting property of this parameterisation is that it results in a constant energy density term mimicking an effective cosmological constant, with $\Omega_\Lambda = -P_0/\left[\rho_c(1+\alpha)\right]$, plus an evolving term that can reproduce a dark matter behaviour for certain choices of the parameter $\alpha$. This allows one to either treat the affine fluid as a single unified dark component, or to use it to model dark energy alone.  

As shown in \cite{quercellini_fields07}, when $\alpha$ is negative, this description can be seen as the attractor solution for a quintessence scalar field dynamics. Alternatively, when treating perturbations, a barotropic affine fluid can be interpreted as a k-essence scalar field (naturally describing an effective cosmological constant plus dark matter), while a scalar field with sound speed $c_s^2=1$ act as a dark energy component. In addition, an affine fluid description can also be interpreted as the result of two interacting dark components (one of them being a cold dark matter component), as we discussed in detail in \cite{QuercelliniEtAl2008_coupled}.
In the present work we explore several different cases resulting from the affine fluid description. We consider two classes of models: one where the affine fluid describes a unified dark component, the other containing a cold dark matter component as well. For each class, we also study three separate subcases, identified by the value of the speed of sound: the barotropic case, with $c^2_{\eff}=\alpha$, the case $c^2_{\eff}=1$, and the ``silent'' case \cite{Bruni:1995b,Bruni:1995a} with $c^2_{\eff}=0$.

To study the properties of the model, we calculate the evolution of scalar perturbations in the affine fluid by modifying the publicly available CAMB code, and set constraints to the parameters of the model by performing  a Monte Carlo Markov Chain analysis using the cosmic microwave background (CMB) anisotropy WMAP 5 year data \cite{WMAP5Komatsu2008} and the large-scale matter distribution derived from the Sloan Digital Sky Survey (SDSS) Luminous Red Galaxy (LRG) 4 year data \cite{sdsslrg4}.

In the next section we describe our model and the theoretical
framework adopted, in Sec.~\ref{results} we discuss the results of our numerical calculations and comparison with observations, and in Sec.~\ref{conclusions} we draw our final conclusions.

%-------------------------------------------------------------------------------------------------------------------------------------
\section{Affine fluid model}

\subsection{General framework}

We perform our calculations in the context of a flat,
homogeneous and isotropic universe, whose unperturbed evolution is described by the Friedman equation
\begin{equation}
H^2\equiv\Big(\frac{\dot{a}}{a}\Big)^2=\frac{8\pi G}{3}\rho\end{equation}
where $\rho$ is the total energy density, sum of the densities of all the components in the universe, each of them satisfying a continuity equation that, in the case of non-interacting components, reads
\begin{equation}
\label{density_eq}
\dot{\rho}_{(i)}+3H(\rho_{(i)}+p_{(i)})=0.
\end{equation}
According to the specific properties of each component one has different scaling  behaviour: for example, for photons and baryons 
$\rho_\gamma\propto a^{-4}$ and $\rho_B\propto a^{-3}$, respectively. We will refer to the decaying in time of the energy density as ``standard'' behaviour; when the energy density grows in time, i.e.\ when $\rho_{(i)}+p_{(i)}<0$ (the null energy condition is violated), the behaviour is called ``phantom'' \cite{2003PhRvL..91g1301C}.

When treating perturbations of the background line element, we adopt the synchronous gauge \cite{MaBertschinger1995}. The perturbed metric then reads:
\begin{equation}
ds^2=a(\tau)^2(d\tau^2-(\delta_{ij}+h_{ij}(\tbf{x},\tau))dx^idx^j)
\end{equation}
where $\tau$ is the conformal time and $\vert h_{ij}\vert \ll 1$ is  the metric perturbation. We then compute the 
Einstein's equations at first order from the
metric given above and from the perturbed energy-momentum tensor
\begin{equation}
T_{\mu\nu}=\sum_i T_{\mu\nu}^{(i)}
\end{equation}
where the index $i$ runs over the components in the universe, photons, baryons, and dark components. The perturbed energy-momentum tensor components are
\begin{eqnarray}
&&T^{(i)0}\,_0=\rho_b^{(i)}(1+\delta^{(i)}),\nonumber\\
&&T^{(i)0}\,_k=\rho_b^{(i)}(1+w^{(i)})V^{(i)}_k,\nonumber\\
&&T^{(i)j}\,_k=(p_b^{(i)}+\delta p^{(i)})\delta^j_k,
\end{eqnarray}
where $\delta^{(i)}$ is the density contrast for the $i$ component, $V^{(i)}$ is the velocity, $w^{(i)}$ is the equation of state parameter (not necessary constant) and the subscript $b$ refers to the background (i.e.\ unperturbed) quantities.

In the next subsections we study in detail the
behaviour of the affine dark component.

\subsection{Background evolution}
The basic property of the phenomenological model we consider is the affine form of the pressure as a function of the density of the dark component, 
Eq.~(\ref{affine_pressure}). Even if the EoS parameter of the  dark component is not constant, a simple solution for the Eq.~(\ref{density_eq}) exists and it is given by
\bea
\label{eq:gen_DF_bg}
\rho_X &=& \rho_\Lambda+(\rho_{X0}-\rho_\Lambda)a^{-3(1+\alpha)}, \qquad \alpha \ne -1; \\
\label{eq:al-1_DF_bg}
\rho_X &=& \rho_{X0} - 3 P_0 \ln a, \qquad \alpha = -1. 
\eea
where $\rho_{X0}$ is the  density of the dark component at the present time (i.e.\ $a=1$) and $\rho_\Lambda\equiv -P_0/(1+\alpha)$, with $\alpha$ and $P_0$ free parameters of the model.  This density evolves in time in a way that can be either standard or phantom, depending on the particular choice of the parameters. A full description of the background properties of such a dark component is given in \cite{LADM07}. Here we want to stress that, in the absence of cold dark matter, this component should both be able to create the gravitational potential necessary to form structures at high redshifts, and to drive the late time acceleration of the universe. With respect to a flat $\Lambda$CDM model, we have an additional degree of freedom, $\alpha$, which is the square of the barotropic sound speed, that allows us to investigate the effective equation of state of the clustering part of the component. 

Since the perturbation equations of the dark component will be written in terms of its equation of state parameter, Eq.~(\ref{w_eq}), it is interesting to explicitly consider the time evolution of $w_X$. 

We first comment on the case $\rho_{X0}-\rho_\Lambda>0$ (Fig.~\ref{fig:DF_w_std}). In this case, if $\alpha>-1$, $w_X(a)$ evolves from the value $\alpha$  approaching the value $-1$; conversely, if $\alpha<-1$, it approaches the value $\alpha$ moving away from $w=-1$. In the former situation, $-3(1+\alpha)<0$ and the dynamical part of the affine component dominates at early times.  When $\alpha<-1$, then $-3(1+\alpha)>0$, so that the evolving dark component increases in time, i.e.\ it has a phantom behaviour, becoming dominant at late times. The slope of the curve obviously depends on $\rho_{X0}$, $P_0$ and $\alpha$.

Let us now consider the case when $\rho_{X0}-\rho_\Lambda<0$ (Fig.~\ref{fig:DF_w_unu}). The behaviour is opposite to the previous case, with the phantom evolution appearing when $\alpha>-1$. In this case there is a divergence of $w$ in the past, making this choice of parameters more problematic. In this paper we will restrict the analysis only to cases with $\rho_{X0}-\rho_\Lambda>0$.

\bfg[!htb]
\incgr[width=\columnwidth]{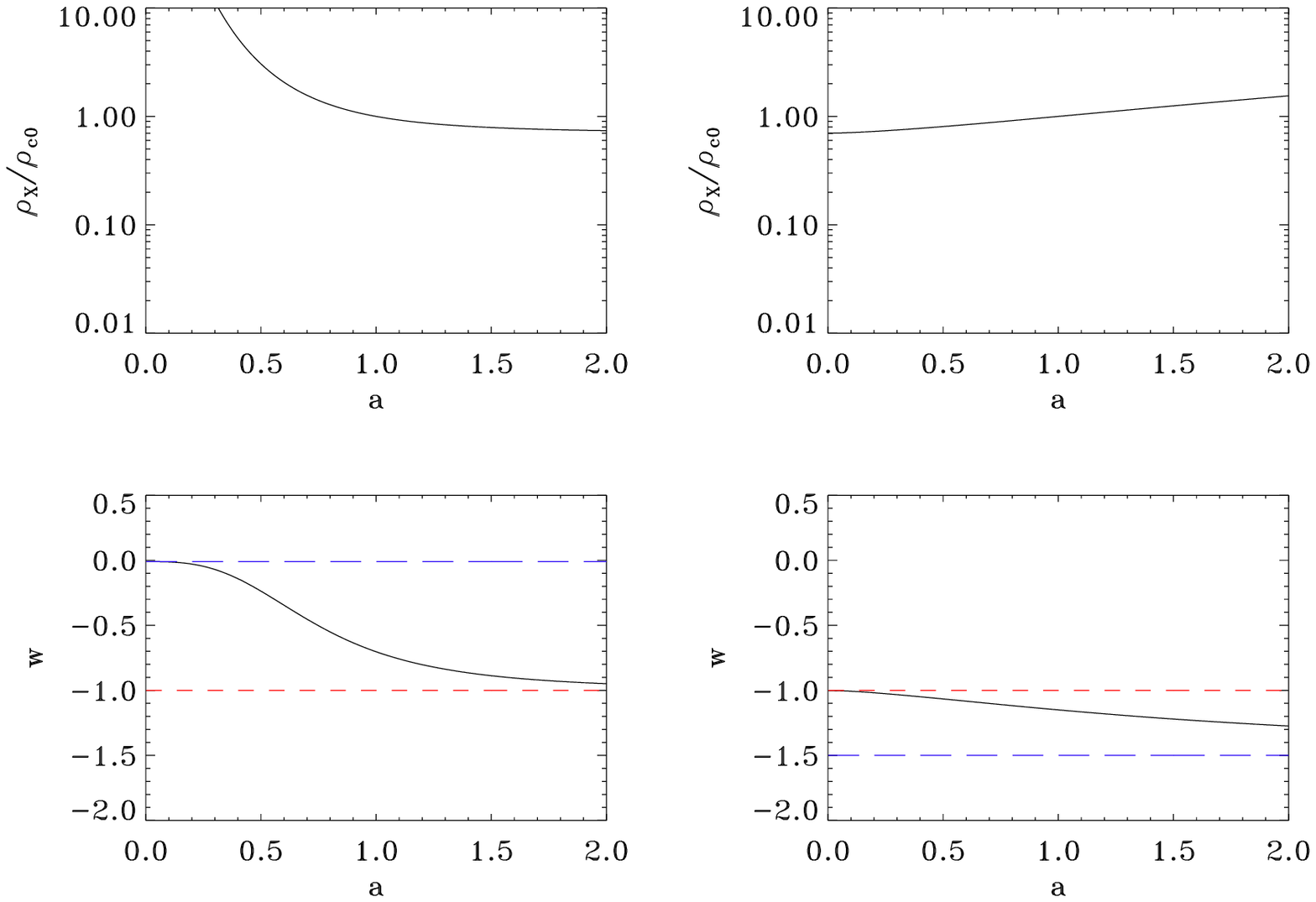}
\caption{Evolution of the dark component energy density (top) and equation of state parameter (bottom), for two values of $\alpha$:  $\alpha=-0.01$ (left) and $\alpha=-1.5$ (right). In both cases, $\rho_{X0}-\rho_\Lambda>0$: in this case, $\alpha<-1$ results in a phantom regime, characterized by an energy density which increases in time. }
\label{fig:DF_w_std}
\efg

\bfg[!htb]
\incgr[width=\columnwidth]{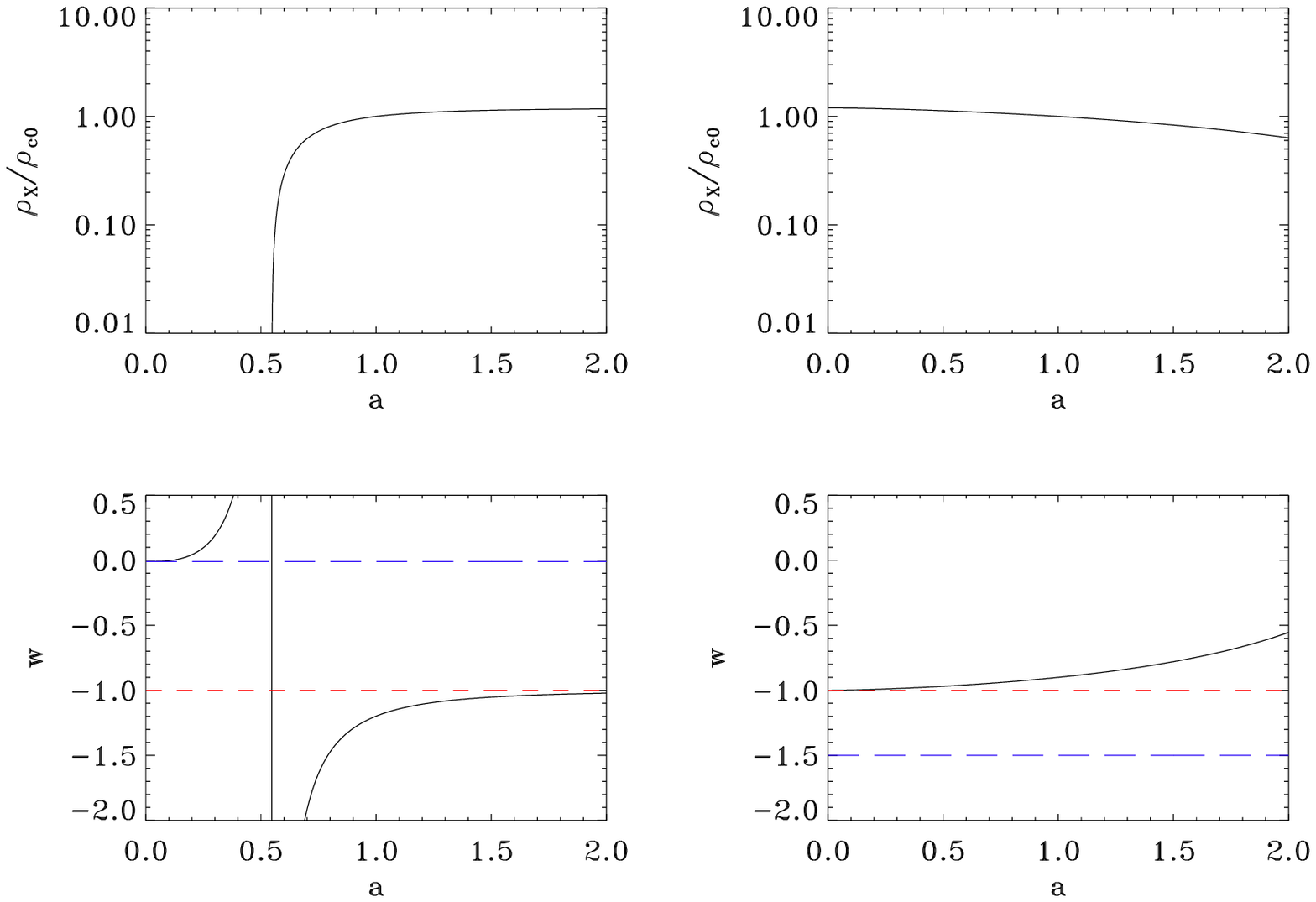}
\caption{Evolution of the dark component energy density (top) and equation of state parameter (bottom): for two values of $\alpha$: $\alpha=-0.01$ (left) and $\alpha=-1.5$ (right). In both cases, $\rho_{X0}-\rho_\Lambda<0$: in this case, $\alpha>-1$ results in a phantom regime, characterized by an energy density which increases in time.} 
\label{fig:DF_w_unu}
\efg

\subsection{Fluid perturbations}

Einstein's equations in the synchronous gauge and in Fourier space give the following system of coupled equations
\begin{eqnarray}
\label{eq:perts_eqs_1}
\dot{\delta}_{(i)}&=&-(1+w_{(i)})\Big(\theta_{(i)}+\frac{\dot{h}}{2}\Big)+\nonumber\\
&+&3H\Big(\frac{dp_{(i)}}{d\rho_{(i)}}-w_{(i)}\Big)\delta_{(i)},\\
\label{eq:perts_eqs_2}
\dot{\theta}_{(i)}&=&-H(1-3w_{(i)})\theta-\frac{\dot{w}_{(i)}}{1+w_{(i)}}\theta_{(i)}+\nonumber\\
&+&\frac{dp_{(i)}/d\rho_{(i)}}{1+w_{(i)}}k^2\delta_{(i)},
\end{eqnarray}
where we defined $ikV_{(i)}\equiv\theta_{(i)}$.

A pure barotropic fluid with a negative EoS parameter has imaginary adiabatic sound speed that causes a runaway growth of perturbations. Not only this has unpleasant consequences for structure formation, but it also creates an instability in the set of coupled perturbation equations (\ref{eq:perts_eqs_1}), (\ref{eq:perts_eqs_2}). A viable way to overcome this inconvenience is to allow for entropy perturbations in the dark component, assuming that the effective speed of sound, sum of the adiabatic and entropic one, is positive or null. We follow the formalism developed in the context of generalised dark  matter \cite{Hu98}, where
\begin{eqnarray}
&& c_{X,\eff}^2\equiv\frac{\delta p_X}{\delta\rho_X}=c_{X,{\rm ad}}^2+\frac{w_{X}}{\delta_{X,{\rm rest}}}\Gamma_X \, ,\\
&& c_{X,{\rm ad}}^2\equiv\frac{\dot p_X}{\dot\rho_X}=\alpha \text{.}
\end{eqnarray}
Here $\Gamma_X$ is a constant parameter we will not use since we prefer to specify the more fundamental quantity $c_{\eff}^2$; $\delta_{X,rest}$ is the density contrast 
in the rest frame of the dark component, defined as
\begin{equation}
\delta_{X,{\rm rest}}=\delta_X+3\frac{\dot a}{a}\frac{\theta_X}{k^2} \text{.}
\end{equation}
The fact that, in our fluid description, the effective speed of sound is a free parameter not tied to the behaviour of equation of state parameter $w_X$, allows us to evade the tight constraints to unified dark matter models pointed out in \cite{2004PhRvD..69l3524S} and arising from the runaway growth of perturbations.

To perform numerical predictions for the evolution of perturbations, we modified the publicly available code CAMB\footnote{http://camb.info/} adding a new component whose perturbations are described by the following equations 
in the synchronous gauge:
\begin{eqnarray}
\label{eq:fluid_eqs}
\dot\delta_X &=& -(1+w_X)(\theta_X+\frac{\dot h}{2})- 3\frac{\dot a}{a}(c_{X,\eff}^2-\alpha)\delta_{X,{\rm rest}}\nonumber\\
&+&\frac{\dot w}{(1+w)}\delta_X,\\
\dot\theta_X &=& -\frac{\dot a}{a}\theta_X+\frac{c_{X,\eff}^2}{(1+w)}k^2\delta_{X,{\rm rest}}.
\end{eqnarray}
We adopt adiabatic initial conditions for the dark component\cite{2003PhRvD..68f3505D, 2004PhRvD..69j3524A}.
We first investigate the constraints coming from the CMB
anisotropy power spectrum on a single dark component governed by an affine equation of state. As we already mentioned, this can account for both dark matter with a non-vanishing EoS parameter and a cosmological constant;
we label this unified model as $\alpha$DM model. The affine component can also be employed as a pure dark 
energy component, if standard CDM is present. We denominate this model as $\alpha$CDM model. In addition to comparing our CMB anisotropy predictions with actual data from the WMAP 5 year observations, we improve our results by adding the SDSS dataset in order to remove degeneracies among  parameters.

In the next section we discuss the results obtained for both $\alpha$DM and $\alpha$CDM models.

%------------------------------------------------------------------------------
\section{Results}
\label{results}

\subsection{Methods}

We performed a full analysis of the two classes of models arising from an affine equation of state (i.e.\ the $\alpha$DM  and $\alpha$CDM models) using the Monte Carlo Markov Chain (MCMC) approach implemented in a modified version of the public CosmoMC software\footnote{http://cosmologist.info/cosmomc/} \cite{LewisBridle2002}. We span the parameter space defined by the baryon density, $\Omega_bh^2$, the cold dark matter density, $\Omega_ch^2$, the current expansion rate of the universe, $H_0$, the reionization optical depth, $\tau$, the spectral index $n_s$ and the normalisation amplitude $A_s$ that parametrise the primordial curvature fluctuation power spectrum 
\begin{equation}
P(k) = A_s (k/k_0)^{n_s}.
\end{equation}
This results in a galaxy power spectrum $P_g(k) = b_L^2 9/25 P(k)$. The affine dark component is characterised by the two parameters $\Omega_\Lambda$ (defined, as usual, as $8\pi G\rho_\Lambda/3 H_0^2$) and $\alpha$. Its effective sound speed squared has been fixed to three different values, namely 0, 1 and $\alpha$, in order to consider the three possible clustering possibilities, namely cold dark matter-like behaviour, scalar field limit and barotropic fluid. We assumed a flat universe and set a Gaussian prior on the Hubble parameter with mean value and standard deviation consistent with the Hubble Space Telescope Key Project, $72\pm8$ km/sec/Mpc \cite{HSTKeyP01}.

We computed the likelihood function of the data using the public code provided by WMAP team\footnote{http://lambda.gsfc.nasa.gov/product/map/dr2/likelihood\_get.cfm}
that includes both the temperature and the polarisation CMB power spectrum (the main effect of the latter being a tighter constraint on the optical depth $\tau$).

Even if at the background level the $\alpha$DM model is equivalent to a dark matter with non-vanishing EoS parameter plus a cosmological constant, there are differences at the perturbation level; moreover, the difference  is conceptual, since the $\alpha$DM model treats the dark sector as a whole, and can even be the result of interacting dark components  \cite{QuercelliniEtAl2008_coupled}. We discuss this class of models in Sec.\ \ref{sec:adm}. The $\alpha$CDM models are discussed in Sec.\ \ref{sec:acdm}. Tables \ref{final_table} and \ref{final_table2} summarise the best fit parameter values for the two classes of models. 
%-----------------------------------------------------------------------------------------------------

\subsection{$\alpha$DM Models}
\label{sec:adm}
In this section we investigate  the properties of a single  dark component described by an affine equation of state. The parameters of this model are $(\Omega_bh^2, \theta, \tau, \ln10^{10}A_s, \Omega_\Lambda, \alpha)$. We expect the model with sound speed $c_{\eff}^2=1$ to be ruled out by the current cosmological datasets: a quintessence scalar field able to drive the late time acceleration of the universe 
expansion prevents structure formation \cite{quercellini_fields07}. We tested our pipeline in the limit of standard $\Lambda$CDM model, i.e. for the
choice $\alpha=0$, obtaining results that are  in excellent agreement with the 5 year WMAP release \cite{DunkleyWMAP52008}.
In the following we describe the results obtained for the three
sub-classes of models we analysed.

\begin{figure}[!h]
\begin{center}
\includegraphics[width=0.85\columnwidth]{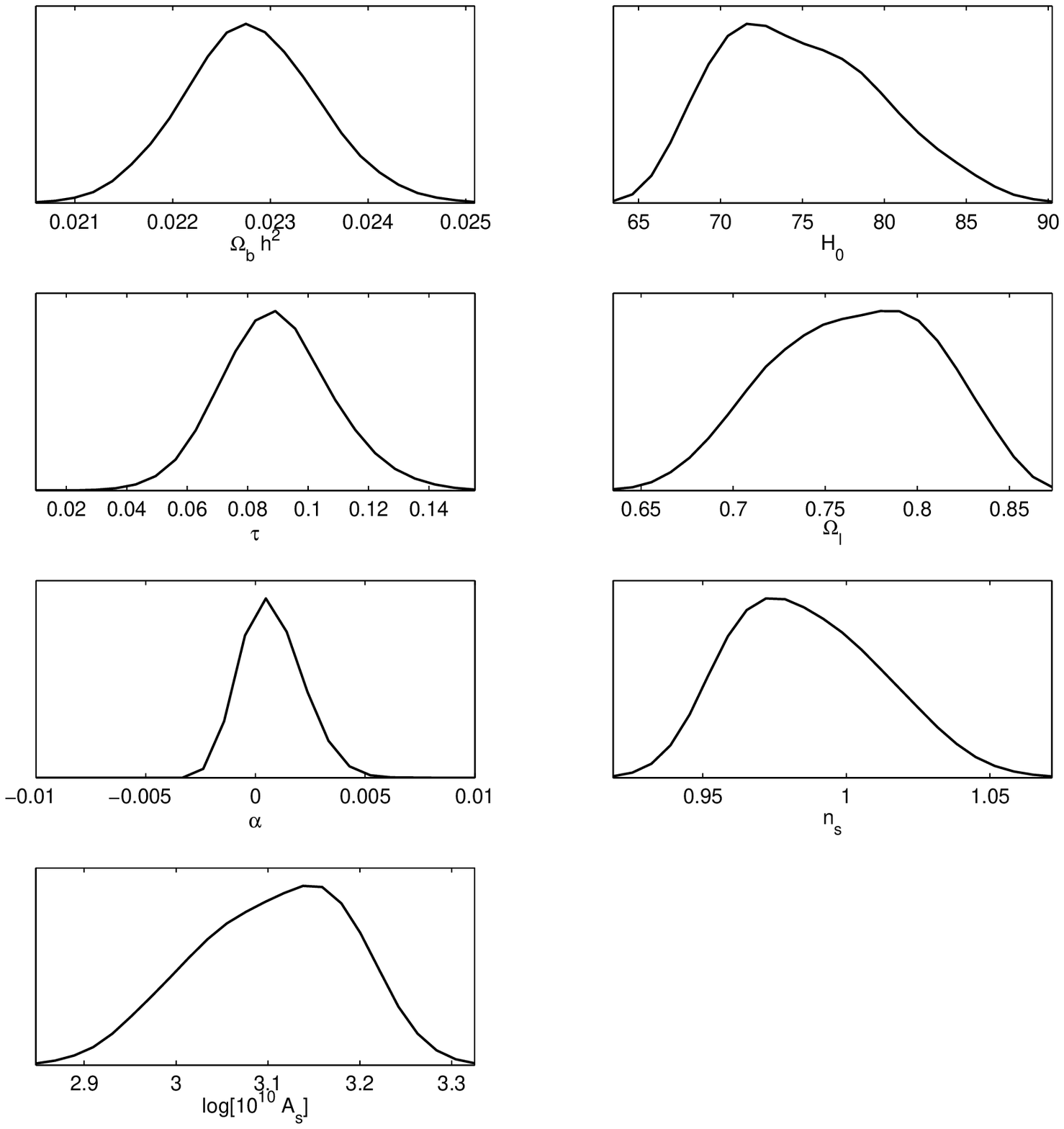}
\includegraphics[width=0.85\columnwidth]{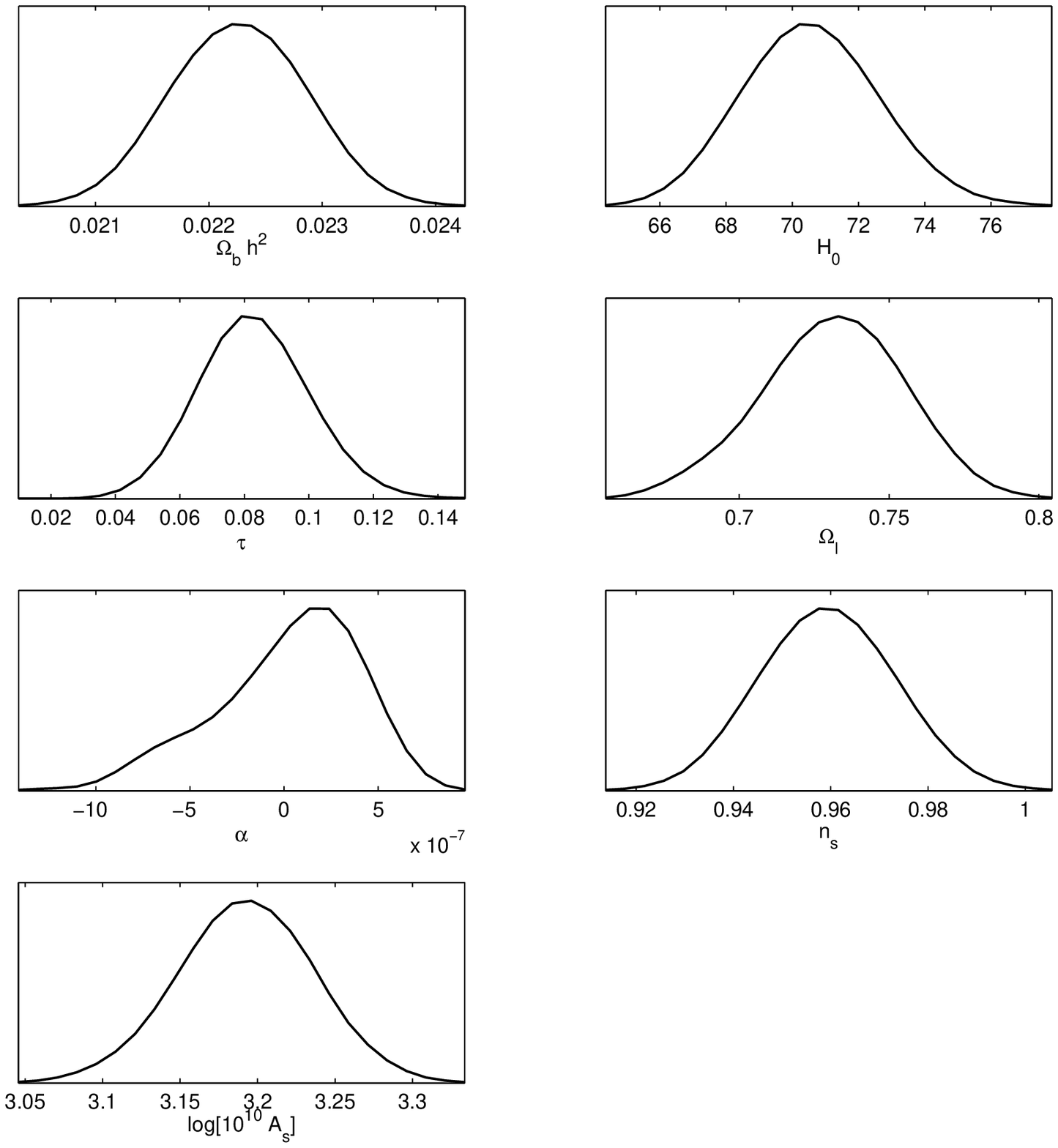}
\caption[barot_likes_mpk]{$\mbf{\alpha}$\tbf{DM -} $\mbf{c_{\eff}^2 = \alpha}$: Parameter likelihoods computed for the $\alpha$DM model under the assumption of barotropic fluid, i.e.\ a fluid that fulfils the relation $c_{\eff}^2=\alpha$. The upper panel is CMB alone, the lower panel is CMB combined with the matter power spectrum. When the matter power spectrum is taken into account the constraints on the equation of state parameter are much tighter. The other parameters are fully consistent with the results of 5 years WMAP release.}
\label{fig:barot_likes}
\end{center}
\end{figure}

%--------------------------------------------
\vspace{0.5 cm}
\paragraph*{$\mbf{\alpha}$\tbf{DM -} $\mbf{c_{\eff}^2 = \alpha}$}
We investigated the barotropic model, namely the one with
$c_{\eff}^2=c_{\rm ad}^2=\alpha$, which 
does not require any assumption concerning entropy perturbations. As we mentioned earlier, this model has an equivalent description in terms of a k-essence scalar field. Our findings are shown in Fig.~\ref{fig:barot_likes}.

\begin{figure}[!h]
\incgr[width=\columnwidth]{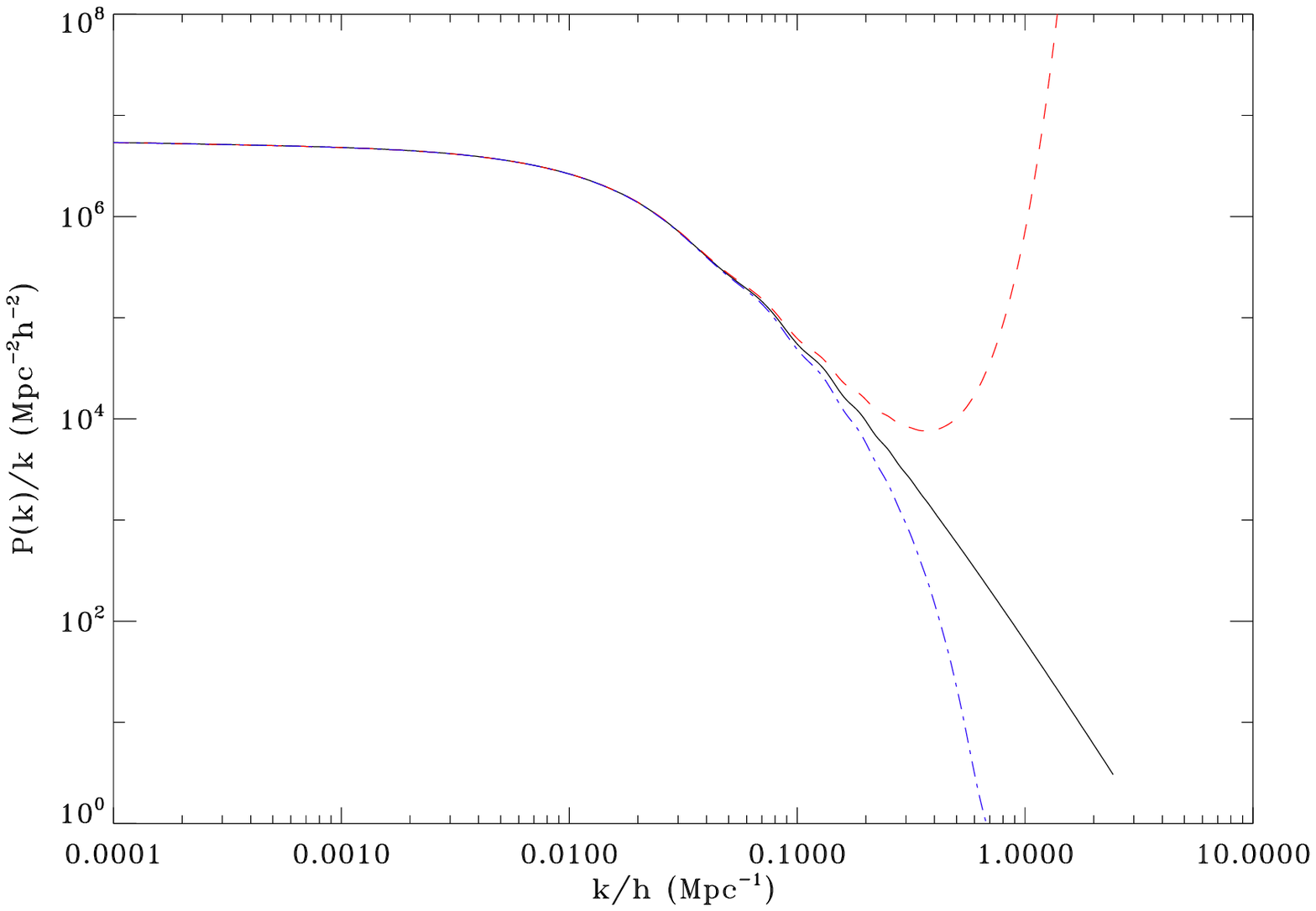}
\caption{Matter power spectrum dependence on $\alpha$. The black solid line is the matter power spectrum computed for  $\alpha=0$, i.e.\ for the concordance $\Lambda$CDM model. The dashed curve is for the value $\alpha=-1\times10^{-6}$; the dot-dashed curve is for  $\alpha=1\times10^{-6}$. The perturbation instability is clear when a negative EoS parameter is  chosen. }
\label{fig:mpsVSalpha1}
\end{figure}

\begin{figure}[!h]
\incgr[width=\columnwidth]{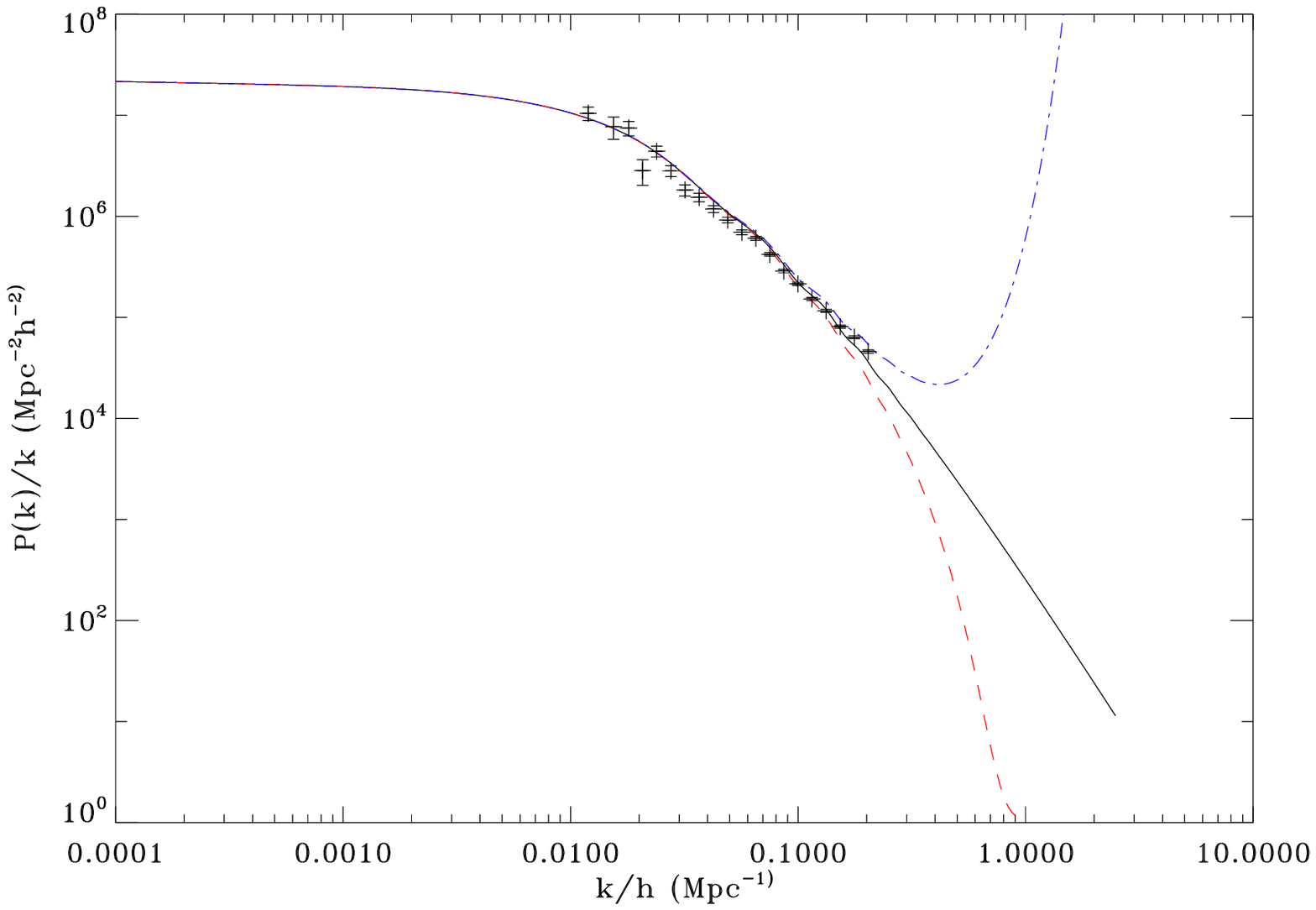}
\caption{To further illustrate the point, for the barotropic $\alpha$DM model we plot against real data the power spectra for values of $\alpha$ at $2\sigma$ from the best fit. It is clear that the data constrain the value of $\alpha$ in two ways: 1) the theoretical curve has to fit the overall  shape of the  data distribution; 2)  the data points at smaller scales  pin down  the value of $\vert \alpha \vert$, constraining it to be small enough to {\it i)} give  a small enough Jeans scale $\lambda_J$ for $\alpha>0$, such that enough power is produced for $\lambda > \lambda_J$,  and {\it ii)}  for $\alpha<0$, to produce an explosive growth of perturbations only at small enough scales, again such that above the Jeans length, where gravity dominates against the pressure effects, the spectrum is undisturbed. It is clear from the figure  that the second effect is dominant, in that  it is extremely sensitive to the value of $\alpha$. }
\label{fig:mpsVSalpha2}
\end{figure}

With this choice of the sound speed we tested the equation of state of dark matter. 
Our best fit model from the 5 year WMAP CMB data has $\alpha=(8\pm11)\times10^{-4}$ and $\Omega_\Lambda = 0.76 \pm 0.04$: we confirm that an almost pressureless component is the most likely one. Since we know that the effect of a non-vanishing sound speed is to strongly modify the clustering properties, we 
investigated the constraints which the matter power spectrum data put on this specific model. As expected, the constraint on $\alpha$ shrinks to $|\alpha|\lesssim10^{-7}$, in excellent agreement with what found in \cite{Muller2004}. For $\Omega_\Lambda$ we find $\Omega_\Lambda = 0.73\pm0.02$. In Figs.~\ref{fig:mpsVSalpha1} and \ref{fig:mpsVSalpha2} the effect of even such a tiny barotropic EoS parameter is shown.

\begin{figure}[!h]
\begin{center}
\includegraphics[width=0.85\columnwidth]{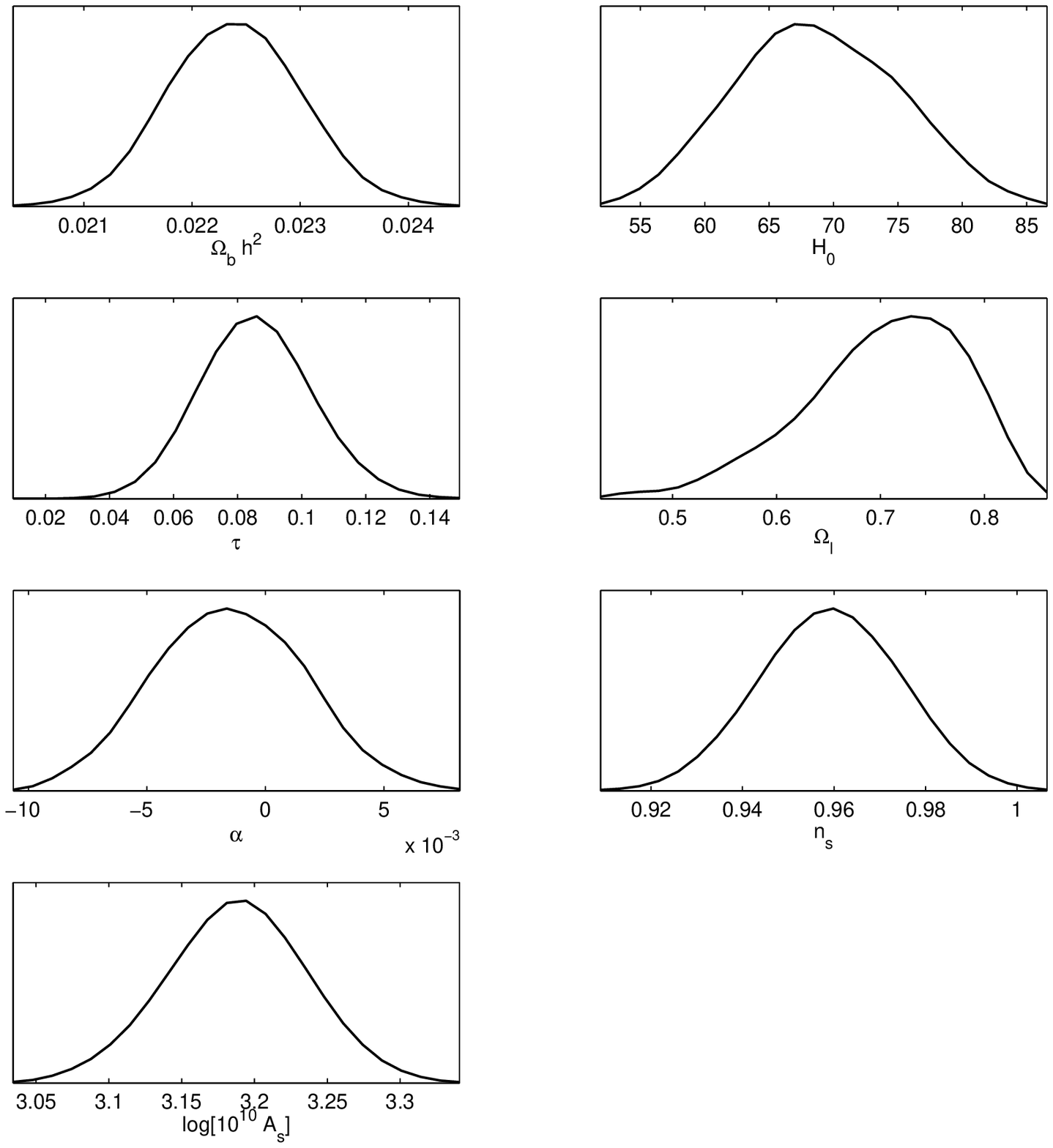}
\includegraphics[width=0.85\columnwidth]{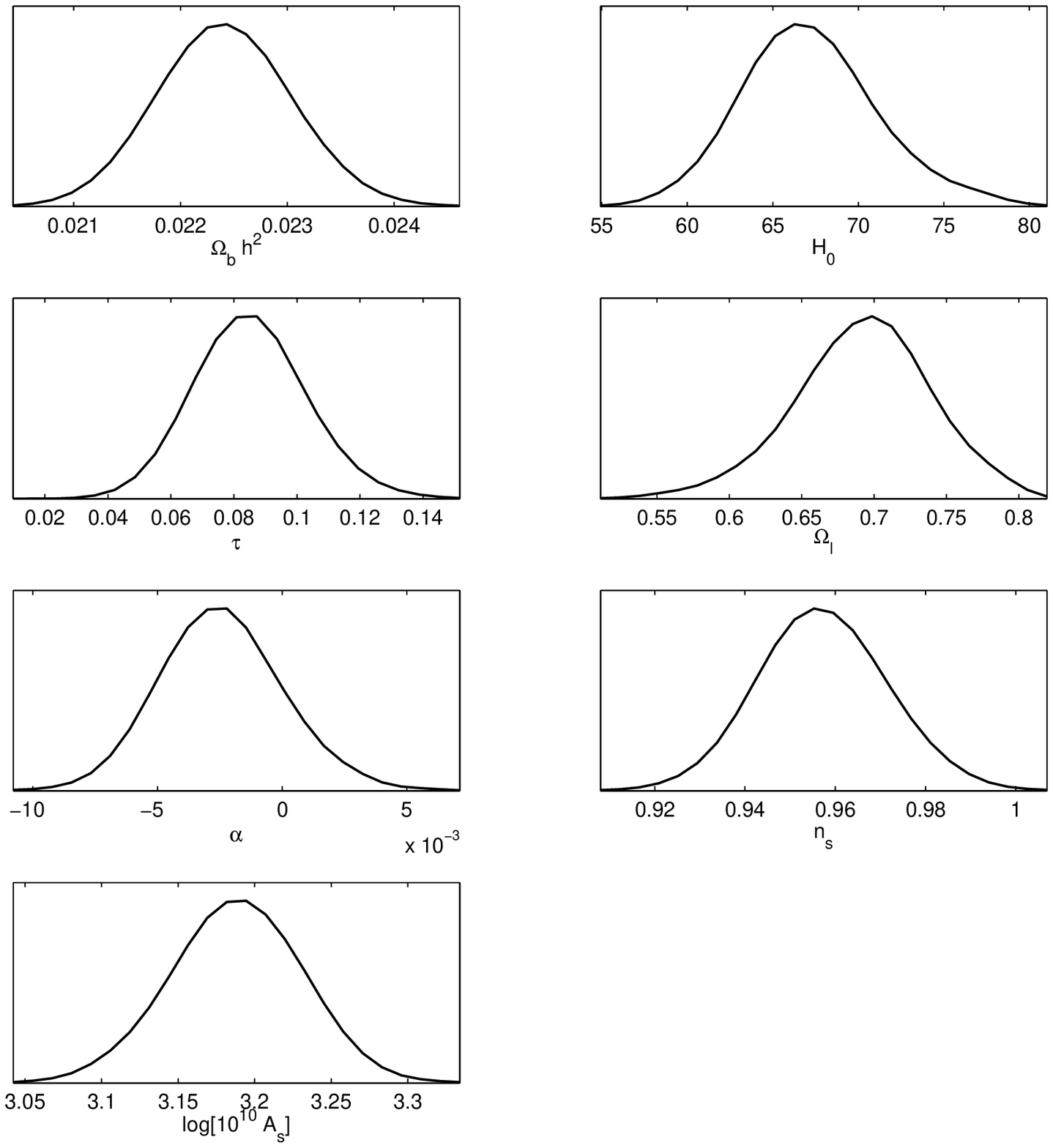}
\caption{$\mbf{\alpha}$\tbf{DM -} $\mbf{c_{\eff}^2 = 0}$:  Parameter likelihoods for the $\alpha$DM model
  with sound speed $c_{\eff}^2=0$. The upper panel is for CMB alone, the lower panel is for CMB combined with the matter power spectrum. The barotropic sound speed squared $\alpha$ is still consistent  with $0$, but the constraints are weaker than in the case of a pure barotropic fluid. The  other parameters 
do not change significantly with respect to the concordance model.}
\label{fig:adm_cs0_likes}
\end{center}
\end{figure}

%--------------------------------------------
\vspace{0.5 cm}
\paragraph*{$\mbf{\alpha}$\tbf{DM -} $\mbf{c_{\eff}^2 = 0}$}
The parameter likelihoods for the case of $c_{\eff}^2=0$ are shown in Fig.~\ref{fig:adm_cs0_likes}. 
The main difference with respect to the barotropic model is a weaker constraint on $\alpha$, due to the presence of a vanishing effective sound speed that cancels the pressure term in the perturbation equations, guaranteeing the clustering properties of the dark component. We get $\alpha=(-1.5\pm 3)\times 10^{-3}$ and $\Omega_\Lambda=0.70\pm 0.09$. When the matter power spectrum is considered, the limit on the square of the barotropic sound speed $\alpha$ shrinks to $(-2\pm 2)\times 10^{-3}$ at 1$\sigma$ level, and $\Omega_\Lambda=0.69\pm 0.05$.

%---------------------------------------------
\vspace{0.5 cm}
\paragraph*{$\mbf{\alpha}$\tbf{DM -} $\mbf{c_{\eff}^2 = 1}$}
For completeness, we also performed the analysis in the weakly clustering limit, described by $c_{\eff}^2=1$; as expected, the model fails completely in fitting the observational data.
A fluid with a luminal speed of sound prevents the clustering at
 scales even close to the horizon \cite{quercellini_fields07}.

%-----------------------------------------------------------------------------------------------------
\subsection{$\alpha$CDM Models}
\label{sec:acdm}
In what follows we present the results we obtained for the $\alpha$CDM model, i.e.\ when we consider a flat universe filled with baryons, cold dark matter  and a dark energy component described by the affine equation of state Eq.\ (\ref{affine_pressure}). The choice can help to
distinguish the cosmological constant from a more general dynamical field. In this framework the most natural value for the speed of sound is $c_{\eff}^2=1$: with this choice, our fluid description represents well the attractor dynamics of a quintessence scalar field, when $\alpha <0$ \cite{quercellini_fields07}.

\begin{figure}[!h]
\begin{center}
\includegraphics[width=.85\columnwidth]{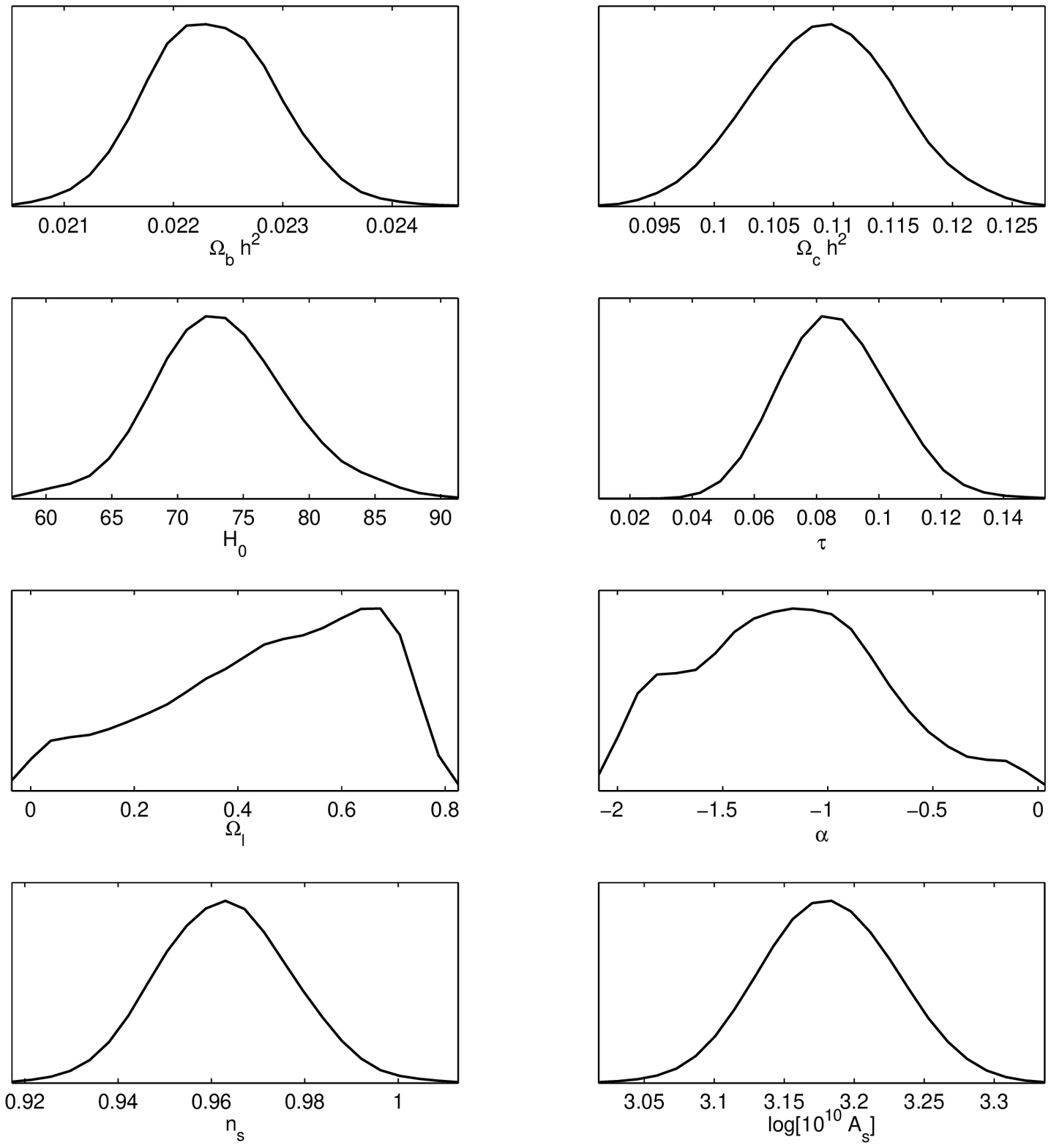}
\includegraphics[width=.85\columnwidth]{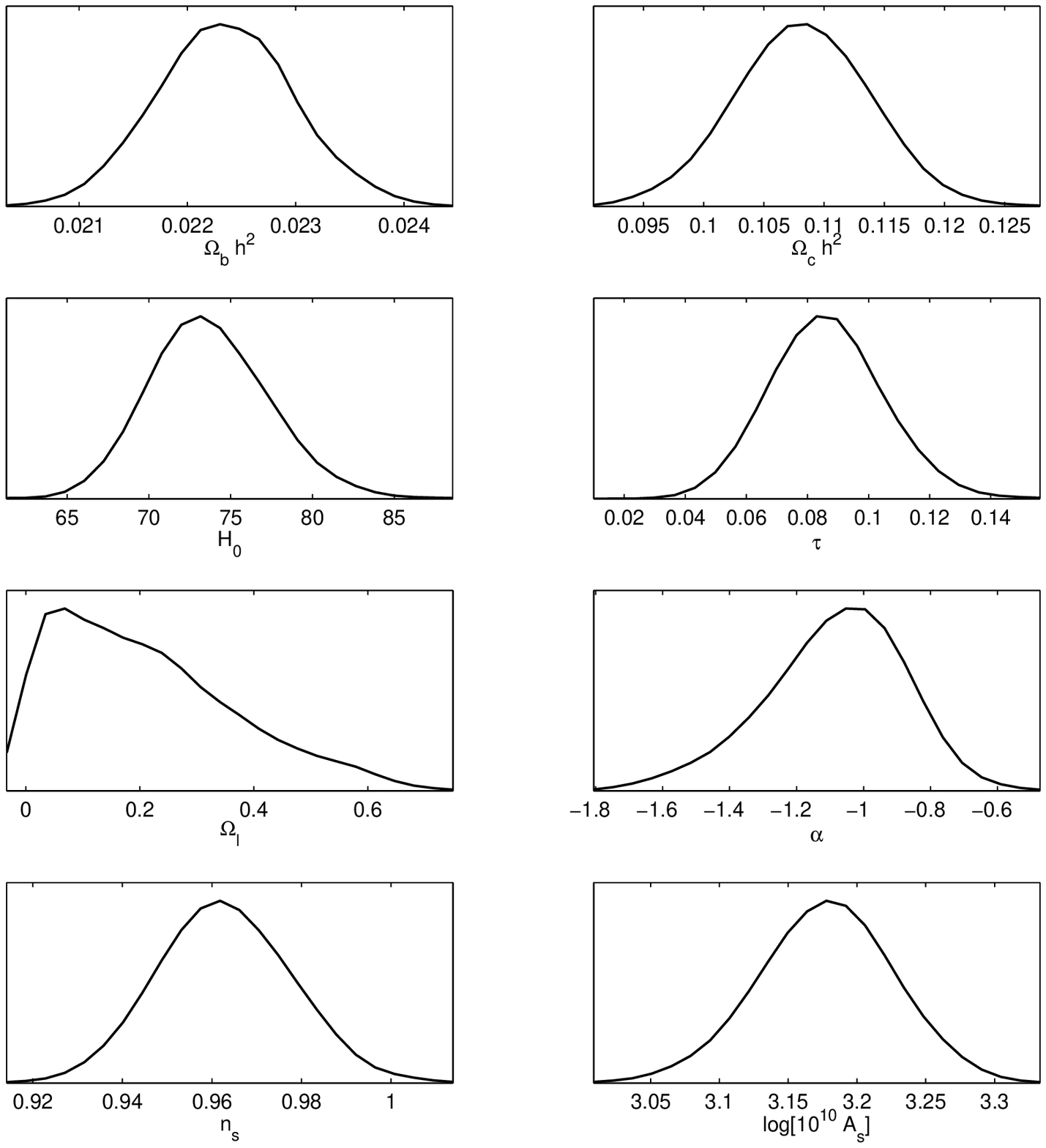}
\caption{$\mbf{\alpha}$\tbf{CDM -} $\mbf{c_{\eff}^2=1}$: Parameter likelihoods computed for  the $\alpha$CDM model when the sound speed is fixed to $c_{\eff}^2 =1$. The upper panel is CMB alone, the lower panel is CMB combined with the 
matter power spectrum. The almost flat likelihood for $\Omega_\Lambda$ together with the broad one for $\alpha$ reflect the degeneracies of the model. Adding matter power spectrum data helps to break this degeneracy since it forces $\Omega_c h^2$ to be of the order of $0.11$ and $\alpha\sim -1$.  However, $\Omega_\Lambda$ remains essentially unconstrained. }
\label{acdm_cs1_likes}
\end{center}
\end{figure}

\vspace{0.5cm}
\paragraph*{$\mbf{\alpha}$\tbf{CDM -} $\mbf{c_{\eff}^2=1}$}
In Fig.~\ref{acdm_cs1_likes} we show the results for the $\alpha$CDM model with $c_{\eff}^2=1$. Also this case, as the previous one, has an equivalent description in terms of a scalar field, but with a  standard kinetic term.  The main effect of dark energy is to modify the low multipoles region of the CMB power spectrum, unfortunately the one where high cosmic variance prevents a precise determination of the cosmological parameters. Even worse, the model suffers of intrinsic degeneracies. At zeroth order, i.e.\ in the background, 1) if $\alpha\sim 0$ the dynamical part of the affine component behaves like dark matter, while, 2) if $\alpha\sim -1$ it can replace the cosmological constant. Since we fixed the speed of sound equal to $1$, the first degeneracy is not present because dark matter and the affine component are different at the perturbation level. We are left with the second degeneracy, that is clearly visible in the flat likelihood for $\Omega_\Lambda$ and the broad likelihood for $\alpha$. We get a rather loose constraint on $\alpha$, i.e.\ $\alpha=-1.2\pm 0.4$, while $\Omega_\Lambda=0.5\pm 0.2$.
When we add the matter power spectrum, the $\Omega_\Lambda-\alpha$ degeneracy is partially removed. The result is a slightly tighter constraint on $\alpha$, which is $\alpha=-1.1\pm 0.2$.

\begin{figure}[!h]
\begin{center}
\includegraphics[width=.85\columnwidth]{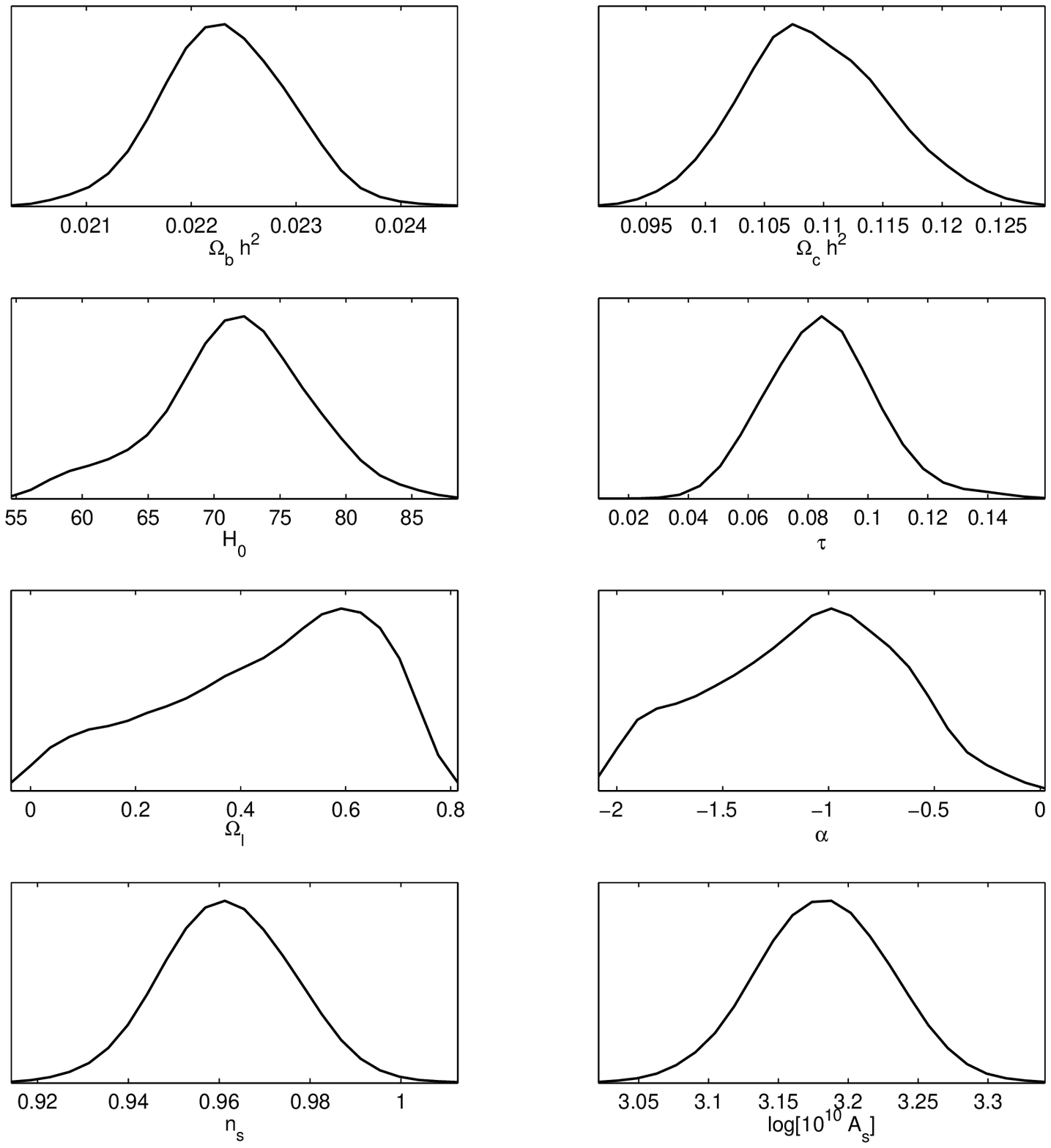}
\includegraphics[width=.85\columnwidth]{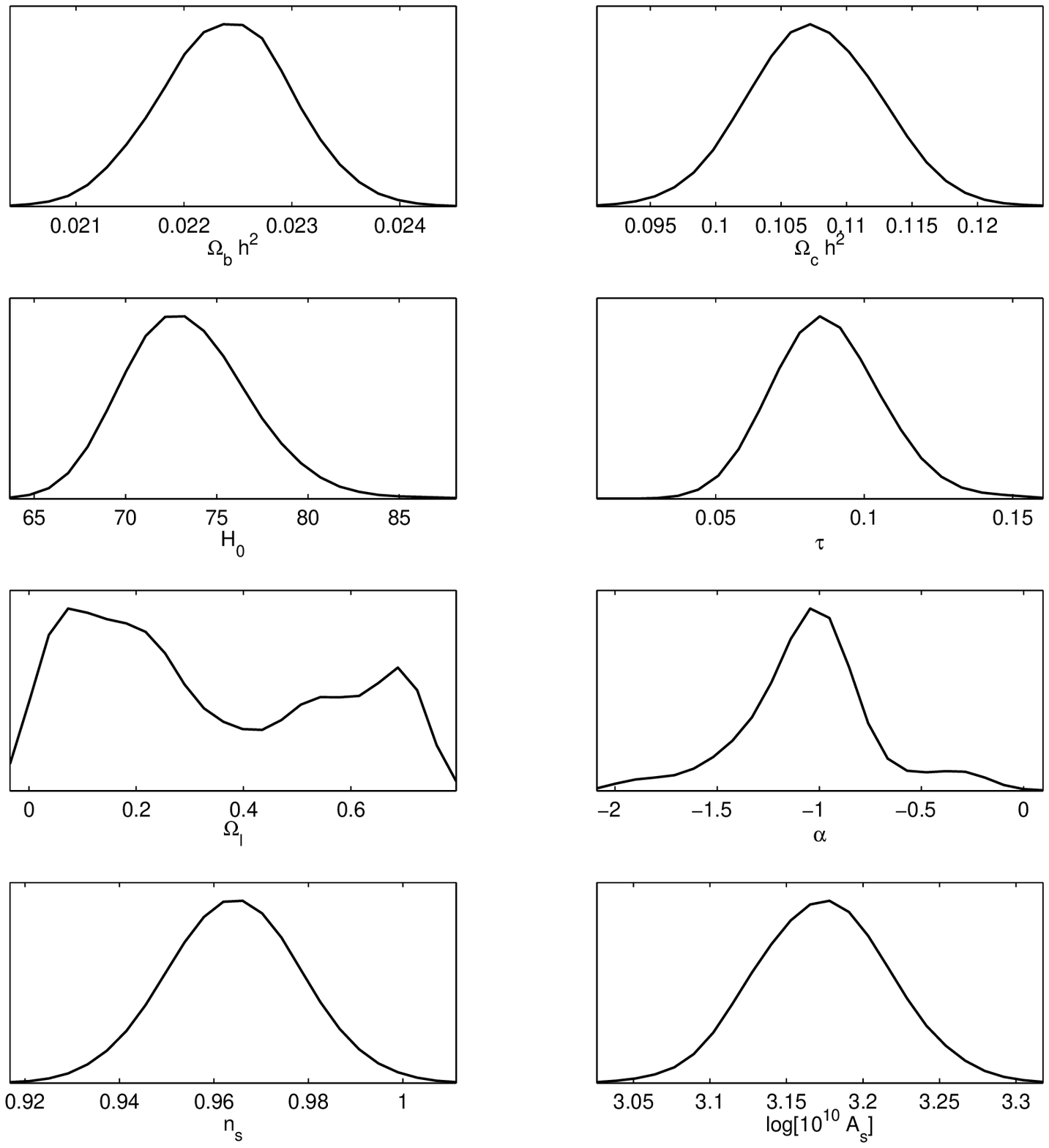}
\caption{$\mbf{\alpha}$\tbf{CDM -} $\mbf{c_{\eff}^2=0}$:  Parameter likelihoods for the $\alpha$CDM model when the
  sound speed is fixed to $c_{\eff}^2=0$. The upper panel is from CMB alone, the lower panel is from CMB combined 
with the matter power spectrum. The results are very close  to those obtained in the case of sound speed equal to 1. }
\label{acdm_cs0_likes}
\end{center}
\end{figure}

\vspace{0.5cm}
\paragraph*{$\mbf{\alpha}$\tbf{CDM -} $\mbf{c_{\eff}^2=0}$}
The results are only marginally affected by the value of the sound speed of the dark component (fig.~\ref{acdm_cs0_likes}), since the CMB is basically insensitive to the sound speed of dark energy.  We find $\alpha=-1.1\pm 0.4$ and $\Omega_\Lambda = 0.5\pm 0.2$. When the matter power spectrum is included in the analysis these change to $\alpha=-1.0\pm 0.3$ and $\Omega_\Lambda = 0.3\pm 0.2$

\begin{figure}[!h]
\begin{center}
\includegraphics[width=.85\columnwidth]{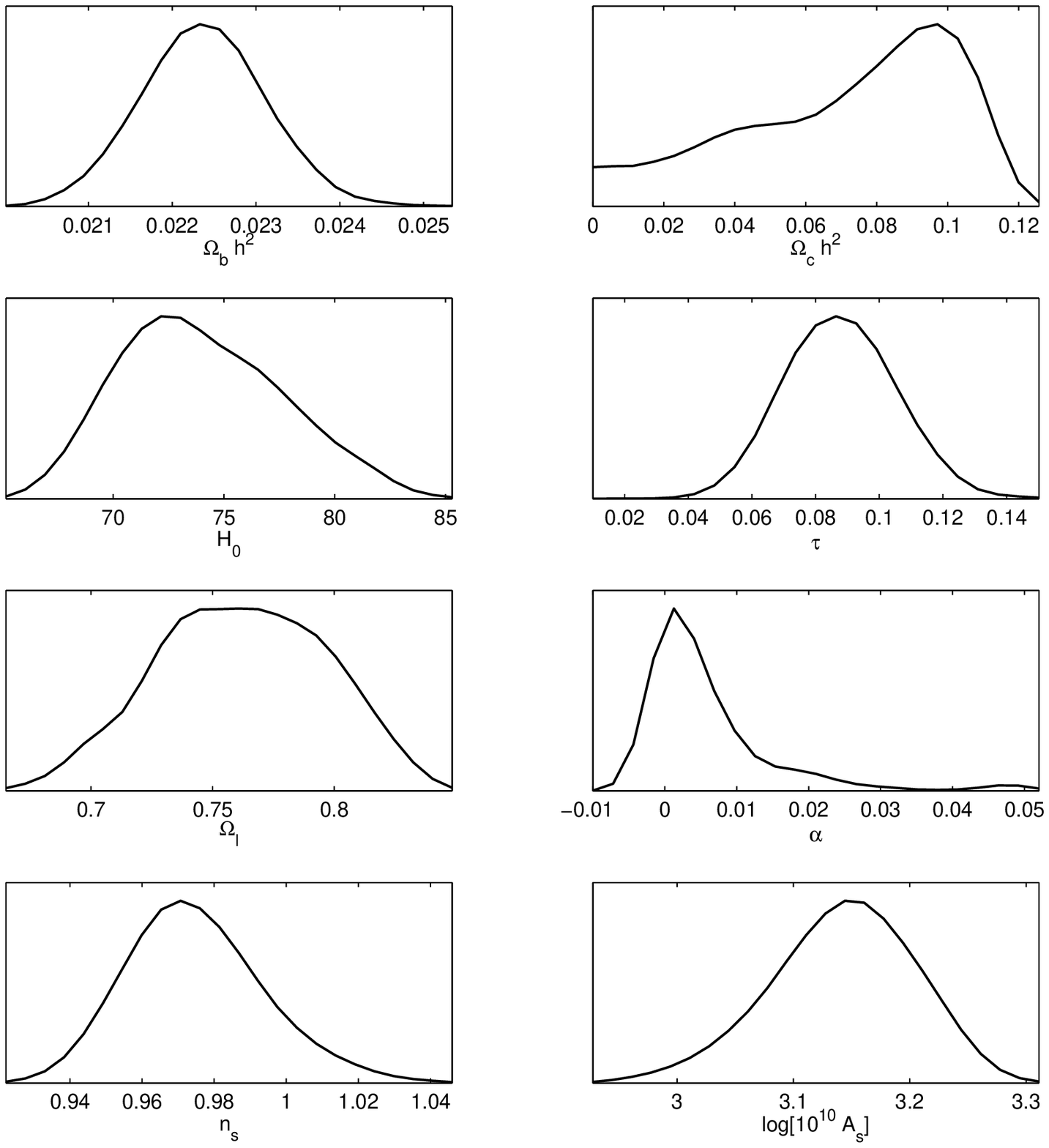}
\includegraphics[width=.85\columnwidth]{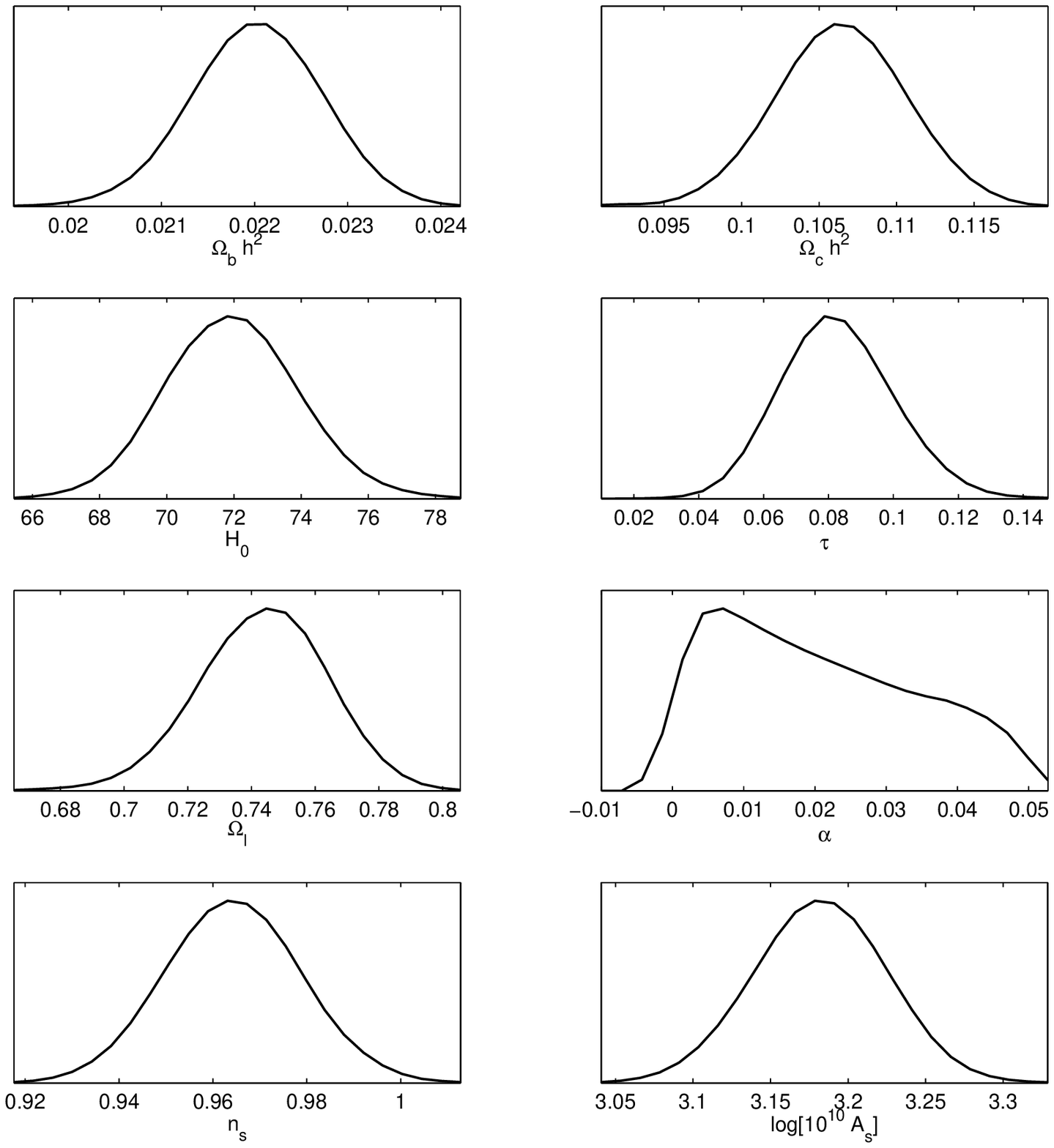}
\caption{$\mbf{\alpha}$\tbf{CDM -} $\mbf{c_{\eff}^2=\alpha}$: CMB alone,
  upper panel, and CMB 
combined with MPS, lower panel, likelihoods for the pure barotropic
  $\alpha$CDM model 
($c_{\eff}^2=\alpha$). The CMB alone likelihoods show the degeneracy
  between $\Omega_ch^2$ 
and $\Omega_\Lambda$, being $\alpha$ close to 1. Adding the matter
  power spectrum 
$\Omega_ch^2$ and $\Omega_\Lambda$ are better constrained, but we lost
  any information 
on $\alpha$ (this is because we assume a flat universe).}
\label{acdm_bar_likes}
\end{center}
\end{figure}

\vspace{0.5cm}
\paragraph*{$\mbf{\alpha}$\tbf{CDM -} $\mbf{c_{\eff}^2=\alpha}$}
When the dark component is forced to be barotropic the only degeneracy we are left with the first degeneracy mentioned above, since $\alpha\sim0$. The result is $\Omega_\Lambda=0.76\pm 0.03$ while $\Omega_c$  and $\alpha$ are badly constrained ($\alpha=(6\pm 9)\times 10^{-3}$): a lower value of $\Omega_c$ can be balanced by the dynamical
 part of the affine component. When the matter power spectrum is added we obtain a slightly tighter constraint on $\Omega_\Lambda$ ($\Omega_\Lambda=0.74\pm 0.02$), while $\Omega_c$ is determined by the shape of the spectrum. This implies actually a broad likelihood for the parameter $\alpha$, since the coefficient $(\rho_{X0}-\rho_\Lambda)\simeq 0$. We obtain $\alpha=(1.9\pm 1.4)\times 10^{-2}$. The figure \ref{acdm_bar_likes} summarises the results described above. 

\section{Conclusions}
\label{conclusions}
We studied the effect of an affine EoS fluid model applied to the dark sector, both as a unified description of dark matter and an effective cosmological constant, and as a pure dark energy component. Our model makes use of a dynamical parameterisation relating $p$ and $\rho$, as opposed to the usual cynematical description of the EoS parameter in terms of its current value and its first derivative. In a previous paper \cite{LADM07} we carried on a comparison of the background evolution of this model with existing cosmological observations. In the present work, we focused on the behaviour of cosmological perturbations, and compared the theoretical predictions with the CMB WMAP 5 year data, and with the SDSS large scale structure data.

\begin{table*}[h!]
\caption[]{Best fit parameter values for $\alpha$DM.}
\label{final_table}
\begin{center}
\begin{tabular}{|c|c|c|c|c|c|}
\hline
models/params  & $\Lambda$CDM +    & \multicolumn{2}{|c|}{$\alpha DM$ - bar}& \multicolumn{2}{|c|}{$\alpha DM$ - $c^2_{\eff}=0$} \\ 
               & SNe \& BAO (WMAP5)  & CMB               &  MPS               & CMB                & MPS               \\ 
\hline
$\Omega_b h^2$ & $0.02265\pm0.00059$ & $0.0223\pm0.0007$ &$0.0223\pm0.0006$ & $0.0224\pm0.0006$  & $0.0224\pm0.0003$ \\ 
\hline
$\Omega_c h^2$ & $0.1143\pm0.0034 $ & -                 & -                  & -                   & -                 \\ 
\hline
$H_0$          & $70.1\pm1.3$       & $75\pm5$          &  $71\pm2$          & $69\pm6$            & $67\pm4$          \\ 
\hline
$\tau$         & $0.084\pm0.016$    & $0.090\pm0.018$   &  $0.083\pm0.016$   & $0.086\pm0.017$     & $0.085\pm0.017$   \\ 
\hline
$n_s$          & $0.960\pm0.014$    & $0.99\pm0.03$     &  $0.960\pm0.014$   & $0.959\pm0.015$     & $0.957\pm0.014$   \\ 
\hline
$\log(10^{10}A_s)|_{k=0.002}$ & $3.20\pm0.08$ & $3.10\pm0.08$ &$3.19\pm0.04$ & $3.19\pm0.05$       & $3.19\pm0.04$     \\ 
\hline
$\Omega_\Lambda$     &$0.721\pm0.015$     & $0.76\pm0.04$  &$0.73\pm0.02$    & $0.70\pm0.09$       & $0.69\pm0.05$     \\ 
\hline
$\alpha$       & -                 & $(8\pm11)\times10^{-4}$& $(0.2\pm4)\times10^{-7}$ & $(-1.5\pm3)\times10^{-3}$ & $(-2\pm2)\times10^{-3}$ \\ \hline
\end{tabular}
\end{center}
\end{table*}

\begin{table*}[h!]
\caption[]{Best fit parameter values for $\alpha$CDM.}
\label{final_table2}
\begin{center}
\begin{tabular}{|c|c|c|c|c|c|c|}
\hline
models/params  &  \multicolumn{2}{c|}{$\alpha CDM$ - bar}& \multicolumn{2}{c|}{$\alpha CDM$ - $c^2_{\eff}=0$} & \multicolumn{2}{c|}{$\alpha CDM$ - $c^2_{\eff}=1$}\\ 
               & CMB                &  MPS               & CMB                & MPS               & CMB                & MPS               \\ 
\hline
$\Omega_b h^2$ &  $0.0224\pm0.0007$ &  $0.0220\pm0.0007$ & $0.0223\pm0.0006$  & $0.0224\pm0.0006$ & $0.0224\pm0.0006$ & $0.0224\pm0.0006$\\ 
\hline
$\Omega_c h^2$ &  $0.07\pm0.03$      &  $0.106\pm0.004$   & $0.109\pm0.006$    & $0.107\pm0.006$   & $0.109\pm0.006$ & $0.107\pm0.006$\\ 
\hline
$H_0$          &  $74\pm4$           &  $72\pm2$          & $71\pm6$           & $73\pm3$          & $73\pm5$ & $74\pm3$\\ 
\hline
$\tau$         &  $0.087\pm0.017$    &  $0.083\pm0.017$   & $0.085\pm0.018$    & $0.088\pm0.017$    & $0.086\pm0.016$ & $0.085\pm0.018$\\ 
\hline
$n_s$          &  $0.975\pm0.019$    &  $0.964\pm0.015$   & $0.962\pm0.014$    & $0.964\pm0.014$   & $0.963\pm0.014$ & $0.963\pm0.015$\\ 
\hline
$\log(10^{10}A_s)|_{k=0.002}$ &  $3.15\pm0.06$ &$3.19\pm0.05$   & $3.18\pm0.05$      & $3.17\pm0.04$     & $3.18\pm0.05$ & $3.17\pm0.05$\\ 
\hline
$\Omega_\Lambda$ &  $0.76\pm0.03$      &  $0.74\pm0.02$     & $0.5\pm0.2$        & $0.3\pm0.2$     & $0.5\pm0.2$ & $0.2\pm0.2$\\ 
\hline
$\alpha$       &  $(6\pm9)\times10^{-3}$ & $(1.9\pm1.4)\times10^{-2}$ & $-1.1\pm0.4$ & $-1.0\pm0.3$      & $-1.2\pm0.4$ & $-1.1\pm0.2$\\ 
\hline
\end{tabular}
\end{center}
\end{table*}

As a first result, we obtained much tighter constraints on the parameters of the model with respect to the analysis carried on the background observables in \cite{LADM07}, confirming that perturbations should be properly included in the calculations when  developing effective models for the dark sector
\cite{ValiviitaMajerottoMaartens2008}. 

In the case when the fluid is treated as a unified dark component, we get values of the effective cosmological constant $\Omega_\Lambda\simeq 0.7$, essentially independent of the speed of sound. For the equation of state parameter $\alpha$, the constraints vary when the fluid is treated as barotropic (resulting in a slightly positive $\alpha$) or a vanishing speed of sound is assumed (resulting in a slightly negative $\alpha$). Both cases are however compatible with $\alpha=0$ at one sigma confidence level. The inclusion of the matter power spectrum in the analysis has generally the effect of shrinking the confidence interval on the parameters, in particular in the barotropic case, due to the effect of $\alpha$ on the Jeans length of the perturbations \cite{quercellini_fields07}.

When standard dark matter is included, the effects of $\alpha$ on the clustering process is less relevant, because the matter-like component of the unified fluid is forced to mimic the cosmological constant behaviour. This is apparent from the fact that the $\alpha$ best fit value moves to $\alpha\sim -1$ which is typical of a cosmological constant. The constraints in the barotropic case remain quite tight, but get larger when the sound speed is set to zero. We also considered the case with a sound of speed equal to unity, which describes a scalar field behaviour. Also in this case the constraints on $\alpha$ are rather loose.

\acknowledgements
DP wishes to thank Giancarlo de Gasperis, Tommaso Giannantonio and Jussi Valiviita for technical support and useful discussions. MB work was partly funded by STFC and partly by a MIUR ``Rientro dei Cervelli" grant.

\bibliography{Balpha}

\end{document}